# Magnetic field dependence of the critical current in a granular s-wave Superconductor

E. Afsaneh and H. Yavari[1]
Department of Physics University of Isfahan, Isfahan, Iran, 81746

**Abstract:** The magnetic field dependences of the critical-current of a granular s-wave superconductor has been determined by considering the rectangular and circular junction model of an array of small superconducting particles which interacting by Josephson coupling through insulating barriers. We will show that in the case of circular model, the maximum critical current of the Josephson current is larger than that of rectangular model.

## 1-Introduction

The physics of small superconducting particles has attracted considerable attention. The behavior of isolated, so-called zero-dimensional (O-D), particles is quite well understood theoretically; at least insofar as their size is not in the microscopic limit [1-3]. The predictions for the magnetic behavior have received a beautiful experimental confirmation [4]. However, our understanding of systems that are composed of small and weakly coupled particles is of a rather preliminary nature. The reason for the interest in such systems is that by varying both the particle size and the interparticle coupling one may span a very wide range of physically interesting situations. In particular, the crossover between 0-D and 3-D (or 2-D for thin films) critical behaviors may be systematically studied. Theoretical studies [5, 6] have emphasized the dependence of the critical behavior on the dimensionality of the system, and the importance of understanding the crossover between different dimensionalities. Furthermore, there is evidence that the systems under consideration can be realized experimentally with grain sizes and inter-grain couplings kept under control [7-11]. One can therefore hope that theoretical predictions may be directly checked by experiment.

A granular superconductor consists of many superconducting islands, the grains. The contacts where the grains touch each other act as weak links interconnecting the grains to form a complex network. Such a system has the properties of a multiply connected superconductor.

The Josephson effect is certainly one of the most intriguing phenomena in superconductivity. It is a consequence of coherent tunneling between two superconducting condensates, each of which is represented by a complex macroscopic wave function, which is the order parameter of superconductivity. On a microscopic level we can describe this effect as the tunneling of Cooper pairs from the pairing state on one side of the junction to that of the other side. In a tunneling process, electrons moving perpendicularly to the interface make the largest contribution. So it follows that the strength of Josephson tunneling will depend on a weighted average over the pairing wave function, weighted in favor of electronic momenta in this perpendicular direction. Therefore the Josephson effect is a direction-sensitive phenomenon connected with the orientation of the junction and with the crystal axis of the superconductor on each side. This fact is of minor importance for conventional s-wave superconductors with an essentially isotropic pair wave function. However, in the case of non-s-wave superconductivity, where pair wave functions have an internal angular structure, this property can lead to intriguing new effects.

Granular superconductors are usually described as a random network of superconducting grains coupled by Josephson weak links [12, 13]. In the high-$T_c$ superconductors (HTCS) ceramics, several experimental groups have found a paramagnetic Meissner effect (PME) at low magnetic fields [14, 15] and they have proposed that this effect could be a consequence of the intrinsic unconventional pairing symmetry of the HTCS of $d_{x^2-y^2}$ type [16]. Depending on the relative orientation of the superconducting grains, it is possible to have weak links with negative Josephson coupling ($\pi$ junctions) [15, 16] which, according to [14, 15], give rise to the PME [17].

[1]Corresponding author: Fax:+983117922409
E-mail address: h.yavary@sci.ui.ac.ir

A significant factor restricting the technical application of high-temperature superconductors (HTSC) is a comparatively low value of the critical transport current density which is mainly determined by their structure like other electromagnetic properties of HTSC. A superconductor prepared by using ceramic technology is a heterogenous system consisting of two phases, viz., granules with a strong superconductivity and a weakly superconducting intergranular phase. The splitting of a superconductor into superconducting regions separated by thin normal layers is a topological effect associated with the structure of oxides and not with the mechanism of high-temperature superconductivity [18].In granular superconductors; the critical current is governed by the electromagnetic coupling across weak intergranular junctions, and the vortex-pinning potential of stronger junctions. Both mechanisms may participate simultaneously in the formation of the critical current. Thus, Gaidukov et al. [19] proposed that the critical current is determined by vortex movement in bulk samples and by the critical current for contacts in films. Using the model of a Josephson medium, Belevtson et al. [20] obtained the field dependence of the critical current density in a superconducting ceramic in the form $J_c(H) = J_c(0)\left(1 - H/H_0\right)$, which is in good agreement with the experimental results.

In this paper, we calculate the critical current of a random network of superconducting grains. To do this we focus on the nature of the Josephson coupling in granular s-wave superconductors. In addition to usual rectangular junction we consider the circular junction. Calculation of the critical current of a granular d-wave superconductor is also under our consideration and will be published elsewhere.

**2- Formalism**

The extremely high sensitivity of the Josephson current to magnetic fields is the key to the most important applications of the Josephson effect. To ensure the critical current density in the current phase relation is independent of the choice of the vector potential, the gauge invariant phase difference is introduced and is defined as [21]

$$\varphi = \varphi_2 - \varphi_1 + \frac{2\pi}{\Phi_0} \int_1^2 \vec{A} \cdot d\vec{l} \tag{1}$$

where $\vec{A}$ is the vector potential and the integration is from the first to the second electrode in a Josephson junction. With this, the general current-phase relation becomes

$$J = J_c \sin\varphi \tag{2}$$

To derive a relation between the gauge invariant phase difference and the magnetic field passing through a junction, consider two pairs of points $Q_1$, $Q_2$ and $P_1$, $P_2$ as in Fig. 1. One has

$$\varphi(P) - \varphi(Q) = \frac{2\pi}{\Phi_0}\left[\int_{P_1}^{P_2} \vec{A}(P) \cdot d\vec{l} - \int_{Q_1}^{Q_2} \vec{A}(Q) \cdot d\vec{l}\right] \tag{3}$$

The magnetic flux through the rectangular contour $\Gamma$ is

$$\Phi = \int_s \vec{B}\,d\vec{s} = \oint \vec{A}\,d\vec{l} = \int_{Q_2}^{Q_1} \vec{A}\,d\vec{l} + \int_{Q_1}^{P_1} \vec{A}\,d\vec{l} + \int_{P_1}^{P_2} \vec{A}\,d\vec{l} + \int_{P_2}^{Q_2} \vec{A}\,d\vec{l} \qquad (4)$$

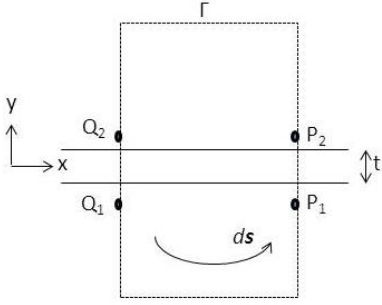

**Figure 1)** The integration contour of the vector potential $\vec{A}$ in a Josephson junction used to derive a relation between the enclosed flux and the phase difference.

If the $Q_1 - P_1$ and $Q_2 - P_2$ portions of the contour are deep enough inside the superconductor, where the current density is essentially zero, the second and fourth term in can be neglected which leads to

$$\varphi(P) - \varphi(Q) = \frac{2\pi}{\Phi_0}\Phi \qquad (5)$$

In general, the gradient of the phase difference along the junction as a function of applied fields can be derived by considering a small section $dx$ of the junction. This leads to

$$\vec{\nabla}\varphi = \frac{2\pi}{\phi_0} d\vec{B} \times \vec{n} \qquad (6)$$

where $\vec{n}$ is a normal vector perpendicular to the junction interface and $d = t + \lambda_R + \lambda_L$ is the (effective) magnetic thickness, with $t$ the barrier thickness and $\lambda_L$ and $\lambda_R$ the (effective) London penetration depth of the respective superconducting electrodes. It is noted here that for thin film electrodes, the penetration depth needs to be corrected due to the finite electrode thickness [22, 23]. With this, the effective magnetic thickness $t$ becomes

$$d = t + \lambda_1 \coth\left(\frac{t_1}{\lambda_1}\right) + \lambda_2 \coth\left(\frac{t_2}{\lambda_2}\right) \qquad (7)$$

Where $t_1$ and $t_2$ are the thickness of the first and second superconducting electrodes, respectively.

As the phase difference along the junction can vary by virtue of the presence of a magnetic field passing through the junction, so it can be the critical currents. The total current which flows through the junction is obtained by integrating the local critical current density $\mathcal{J}(x)$ over the junction area $A$.

In general, this can be written in the form

$$I(B) = \left|\int_{-\infty}^{\infty} dx\, \mathcal{J}(x)\, e^{i\eta x}\right| \qquad (8)$$

Where $\mathcal{J}(x) = \int dy\, j_1(x,y)$ ($j_1$ is the critical density of the Josephson current in the junction) and $\eta = 2\pi B d/\phi_0$.

We consider two specific junction geometries: rectangular and circular for s-s junctions (Fig. 2).

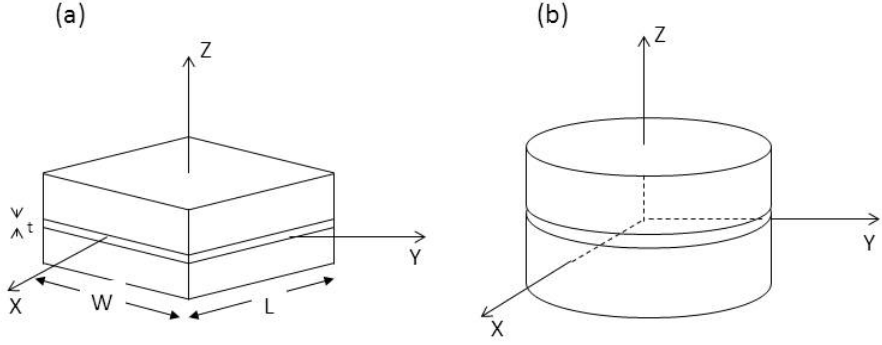

**Figure 2)** Geometrical configuration: (a) Rectangular geometry (b) Circular geometry

In the case of a rectangular Josephson junction with dimensions much smaller than a few times the Josephson penetration depth, the fields from the tunneling currents can be neglected and Eq. (8) can be explicitly integrated, with the result

$$I_{Rec}\left(\frac{\phi}{\phi_0}\right) = I_1 \left|\frac{\sin(\pi\frac{\phi}{\phi_0})}{\pi\frac{\phi}{\phi_0}}\right| \qquad (9)$$

where $\phi = BLd$, is the magnetic flux through the junction, $I_1 = J_1 WL$, and $\eta = \frac{2\pi Bd}{\phi_0}$ ($\phi_0 = \frac{hc}{2e}$). This expression resembles a magnetic field dependence of the critical current pattern as shown in Fig. 3 (a), which is the hallmark of a rectangular junction with homogeneous critical current densities, and is well-known as the Fraunhofer pattern.

For circular junction one has

$$J(x) = \int_{-\sqrt{R^2-x^2}}^{\sqrt{R^2-x^2}} dy\, J_1 = 2J_1\sqrt{R^2 - x^2} \qquad (10)$$

where $R$ is the radius of the junction.

The maximum Josephson current is

$$I_1(k) = \left|2J_1 \int_{-R}^{R} dx\, \sqrt{R^2 - x^2}\, e^{i\eta x}\right| \qquad (11)$$

Doing the integral in Eq. (11) one has

$$I_{Cir}\left(\frac{\phi}{\phi_0}\right) = 2I_1 \left|\frac{K_1(\pi\frac{\phi}{\phi_0})}{\pi\frac{\phi}{\phi_0}}\right| \qquad (12)$$

where $I_1 = \pi R^2 j_1$, $K_1(x)$ is a Bessel function of the first kind, and $\phi = 2BRd$ (Fig. 3 (b)).

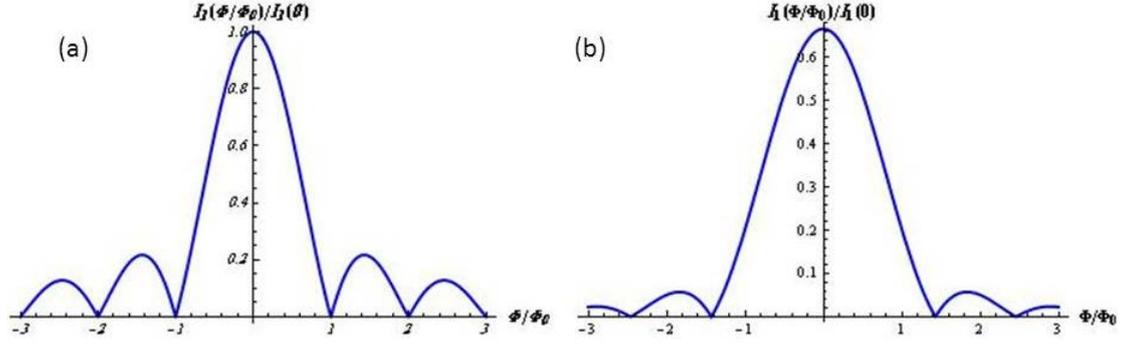

**Figure 3)** Theoretical magnetic field dependence of the maximum Josephson current in s-s junction with (a) Rectangular (b) Circular geometry

We consider Meilikhov's model of deformed granules [21], the weak link between which is formed in the region of plane segments. In this case, the banks of an intergranular Josephson junction are in the form of a circle whose radius $r$ is proportional to the granule size $r = ka$ ($a$ is the average granule size). Suppose that $d$ is the junction thickness Eqs. (9) and (12) respectively reduce to

$$I_{Rec}(B_0) = j_1 \pi r^2 \frac{\sin(\pi \frac{rdB_0}{\phi_0})}{\pi \frac{rdB_0}{\phi_0}} \qquad (13)$$

$$I_{Cir}(B_0) = j_1 \pi r^2 \left( \frac{K_1(\pi \frac{rdB_0}{\phi_0})}{\frac{\pi rdB_0}{2 \phi_0}} \right) \qquad (14)$$

We stress that the magnetic field behavior of the critical current is strongly depends on the form of the grain distribution function. For performing the averaging, we restrict ourselves to Gaussian distribution law. Since it is quite difficult to average the modulus in Eqs. (13) and (14), we calculate the mean square critical current of the junctions. If the randomness of the Josephson lattice is governed by Gauss-like fluctuations of the form

$$P(r) = \frac{32 r^2}{\pi^2 k^3 a^3} e^{-\frac{4r^2}{\pi k^2 a^2}} \qquad (15)$$

The configurational averaging in Eqs. (13) and (14) leads to following equations

$$<I_{Rec}^2(H_a)> = \frac{\pi k^2 a^2 j_1^2 \phi_0^2}{16 d^2 H_a^2} \left\{ 3 - \left[ 3 - 4 \frac{\pi^3 k^2 a^2 d^2 H_a^2}{4\phi_0^2} \left( 3 - \frac{\pi^3 k^2 a^2 d^2 H_a^2}{4\phi_0^2} \right) \right] e^{-\frac{\pi^3 k^2 a^2 d^2 H_a^2}{4\phi_0^2}} \right\} \qquad (16)$$

$$<I_{Cir}^2(H_a)> = \frac{\pi j_1^2 k^2 a^2 \phi_0^2}{2 d^2 H_a^2} \left\{ \left( \frac{4\phi_0^2}{\pi^3 k^2 a^2 d^2 H_a^2} \right) \left[ 1 + 2 \left( \frac{4\phi_0^2}{\pi^3 k^2 a^2 d^2 H_a^2} \right) \right] - \left[ 2 + 3 \left( \frac{4\phi_0^2}{\pi^3 k^2 a^2 d^2 H_a^2} \right) + 2 \left( \frac{4\phi_0^2}{\pi^3 k^2 a^2 d^2 H_a^2} \right)^2 \right] e^{-\frac{\pi^3 k^2 a^2 d^2 H_a^2}{4\phi_0^2}} \right\} \qquad (17)$$

The mean square critical current of granular systems behave smoother than one junction system(Fig 3), which is because of Gaussian average in its critical current where is shown in Fig 4 (a) and (b) for rectangular and circular model respectively.

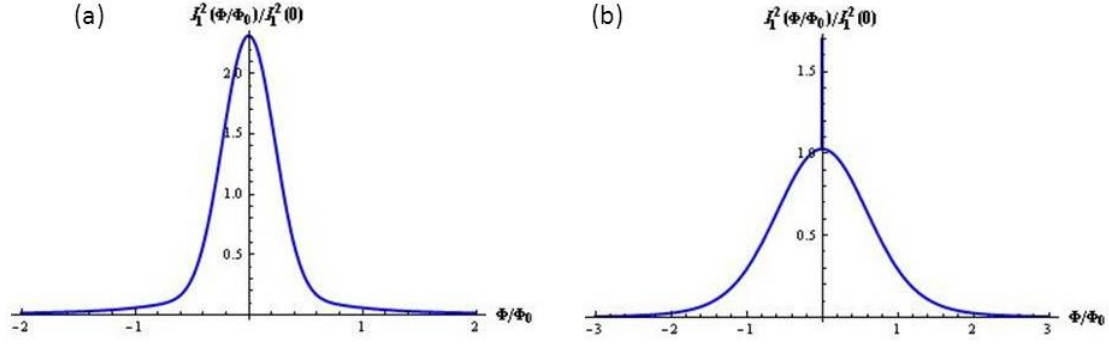

**Figure 4)** Mean square critical current of granular systems with s-s junction in: (a) Rectangular (b) Circular geometry

Figure 5 shows the comparison between the Rectangular and Circular models.

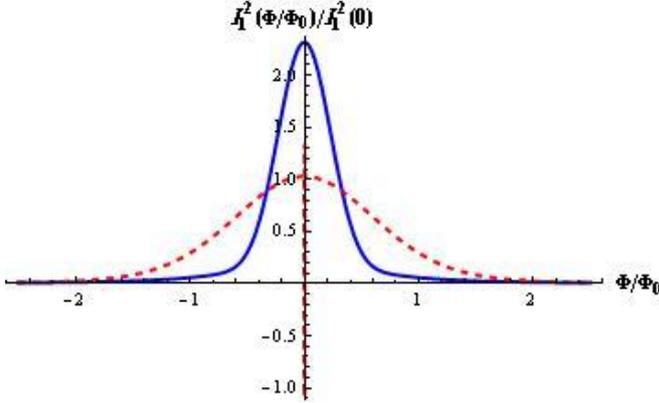

**Figure 5)** Comparison with mean square critical current of granular system s-s junction with "solid line, rectangular model" and "dashed line, circular model"

Eqs. (16) and (17) can be approximated as

$$<I_{Rec}^2(H_a)> = \frac{I_{Max}^2}{1+\frac{H_a^2}{H_J^2}} \tag{18}$$

$$<I_{Cir}^2(H_a)> = \frac{16}{15} I_{Max}^2 \left(1 + \frac{8}{15}\frac{H_J^2}{H_a^2}\right) \tag{19}$$

Where $I_{Max} = \frac{\sqrt{15}}{8}\pi^2 k^2 a^2 j_1$ is the maximum critical current for rectangular case $(H_a = 0)$ and $H_J = \frac{\sqrt{15}\phi_0}{2\pi^{3/2} kad}$.

For circular junction the maximum current $(H_a = 0)$ is

$$<I_{Cir}^2(0)> = \frac{\pi^4}{4} j_1^2 k^4 a^4 = \frac{16}{15} I_{Max}^2 \tag{20}$$

Thus in the circular model, the magnitude of maximum mean square critical current is larger than that of rectangular model by factor $\frac{16}{15}$.

## 3-Conclusions

We have discussed the dynamics of Josephson coupling between superconducting grains in granular s-wave superconductors. In addition to the usual rectangular junction we have also considered the circular junction model. As the magnetic field behavior of the critical current is strongly depends on the form of the grain distribution function, the calculation of mean square critical current of the junctions, leads us to Gaussian distribution law. Our calculations show that the maximum critical-current of a granular superconductor with circular model is larger than that of rectangular model.

**4-References**